\documentclass[a4paper,fleqn]{cas-dc} % twocolumn
\usepackage[numbers]{natbib}

\usepackage{amsfonts}
\usepackage{amsmath}
\usepackage[utf8]{inputenc}

\usepackage{subfig}
\graphicspath{%
{images_paper/}%
}

\usepackage{booktabs}

\begin{document}

\let\WriteBookmarks\relax
\def\floatpagepagefraction{1}
\def\textpagefraction{.001}

\shorttitle{Exploration of Overlap Volumes for Radiotherapy Plan Evaluation with the Aim of Healthy Tissue Sparing}

\shortauthors{M Schlachter et~al.}

\title [mode = title]{Exploration of Overlap Volumes for Radiotherapy Plan Evaluation with the Aim of Healthy Tissue Sparing}
%\tnotemark[1,2]

% \tnotetext[1]{This document is the results of the research
%    project funded by the National Science Foundation.}

% \tnotetext[2]{The second title footnote which is a longer text matter
%    to fill through the whole text width and overflow into
%    another line in the footnotes area of the first page.}

\author[1]{Matthias Schlachter}%
[
%type=editor,
% auid=000,bioid=1,
% prefix=Sir,
% role=Researcher,
orcid=0000-0003-0732-2785]

\cormark[1]


%\ead{cvr_1@tug.org.in}

%\ead[url]{www.cvr.cc, cvr@sayahna.org}

%\credit{Conceptualization of this study, Methodology, Software}

\affiliation[1]{organization={VRVis Zentrum für Virtual Reality und Visualisierung Forschungs-GmbH},
    %addressline={Radarweg 29},
    city={Vienna},
    % citysep={}, % Uncomment if no comma needed between city and postcode
    %postcode={1043 NX},
    country={Austria}}

\author[2]{Samuel Peters}

\author[2]{Daniel Camenisch}
% \ead{cvr3@sayahna.org}
% \ead[URL]{www.sayahna.org}

%\credit{Data curation, Writing - Original draft preparation}

\affiliation[2]{department={Department of Radiation Oncology},
    organization={Kantonsspital St. Gallen},
    %addressline={Mepukada},
    city={St. Gallen},
    % citysep={}, % Uncomment if no comma needed between city and postcode
    %postcode={695571},
    %state={Trivandrum},
    country={Switzerland}}
\author[2,3]{Paul Martin Putora}
% \cormark[2]
% \fnmark[1,3]
% \ead{rishi@stmdocs.in}
% \ead[URL]{www.stmdocs.in}

\affiliation[3]{department={Department of Radiation Oncology},
    organization={University of Bern},
    %addressline={Mepukada},
    city={Bern},
    % citysep={}, % Uncomment if no comma needed between city and postcode
    %postcode={695571},
    %state={Trivandrum},
    country={Switzerland}}

% fith author
\author[1]{Katja B\"uhler}

\cortext[cor1]{Corresponding author}
% \cortext[cor2]{Principal corresponding author}

% % Footnote text
% \fntext[fn1]{This is the first author footnote. but is common to third
% \fntext[fn2]{Another author footnote, this is a very long footnote and
%   it should be a really long footnote. But this footnote is not yet
%   sufficiently long enough to make two lines of footnote text.}

% % For a title note without a number/mark
% \nonumnote{This note has no numbers. In this work we demonstrate $a_b$
%   the formation Y\_1 of a new type of polariton on the interface
%   between a cuprous oxide slab and a polystyrene micro-sphere placed

\begin{abstract}
    \noindent\textbf{Purpose:} Development of a novel interactive visualization approach for the exploration of radiotherapy treatment plans with a focus on overlap volumes with the aim of healthy tissue sparing.

    \noindent\textbf{Methods:}
    We propose a visualization approach to include overlap volumes in the radiotherapy treatment plan evaluation process. Quantitative properties can be interactively explored to identify critical regions and used to steer the visualization for a detailed inspection of candidates. We evaluated our approach with a user study covering the individual visualizations and their interactions regarding helpfulness, comprehensibility, intuitiveness, decision-making and speed.

    \noindent\textbf{Results:} A user study with three domain experts was conducted using our software and evaluating five data sets each representing a different type of cancer and location by performing a set of tasks and filling out a questionnaire. The results show that the visualizations and interactions help to identify and evaluate overlap volumes according to their physical and dose properties. Furthermore, the task of finding dose hot spots can also benefit from our approach.

    \noindent\textbf{Conclusions:}
    The results indicate the potential to enhance the current treatment plan evaluation process in terms of healthy tissue sparing.
\end{abstract}

% % Use if graphical abstract is present
%     \includegraphics[width=\linewidth]{interaction_overview3.pdf}

% % Research highlights
% \begin{highlights}
%     \item Interactive visualization for the exploration of radiotherapy treatment plans
%     \item Focus is on overlap volumes with the aim of healthy tissue sparing
%     \item Quantitative properties can be interactively explored to identify critical regions
%     \item Interaction is used to steer the visualization for a detailed inspection
%     \item Results indicate the potential to enhance the treatment plan evaluation process
% \end{highlights}

% Keywords

\ExplSyntaxOn
\keys_set:nn { stm / mktitle } { nologo }
\ExplSyntaxOff

\maketitle

\section{Introduction}
\label{sec:introduction_overlap}

Radiotherapy (RT) uses high doses of radiation, which can cause tissue damage and severe harm to nearby organs, and may even lead to secondary cancer~\cite{Bentzen2010}.
While RT technology has advanced in recent years, there are still many challenges to overcome~\cite{Schlachter2019a}, such as ensuring minimal damage to normal tissue near a tumor~\cite{Moore2011}. This is a difficult task as it requires balancing the need to control the tumor while also protecting surrounding organs~\cite{Ge2019}.

\begin{figure}[tb]
    \centering

    \includegraphics[width=\linewidth]{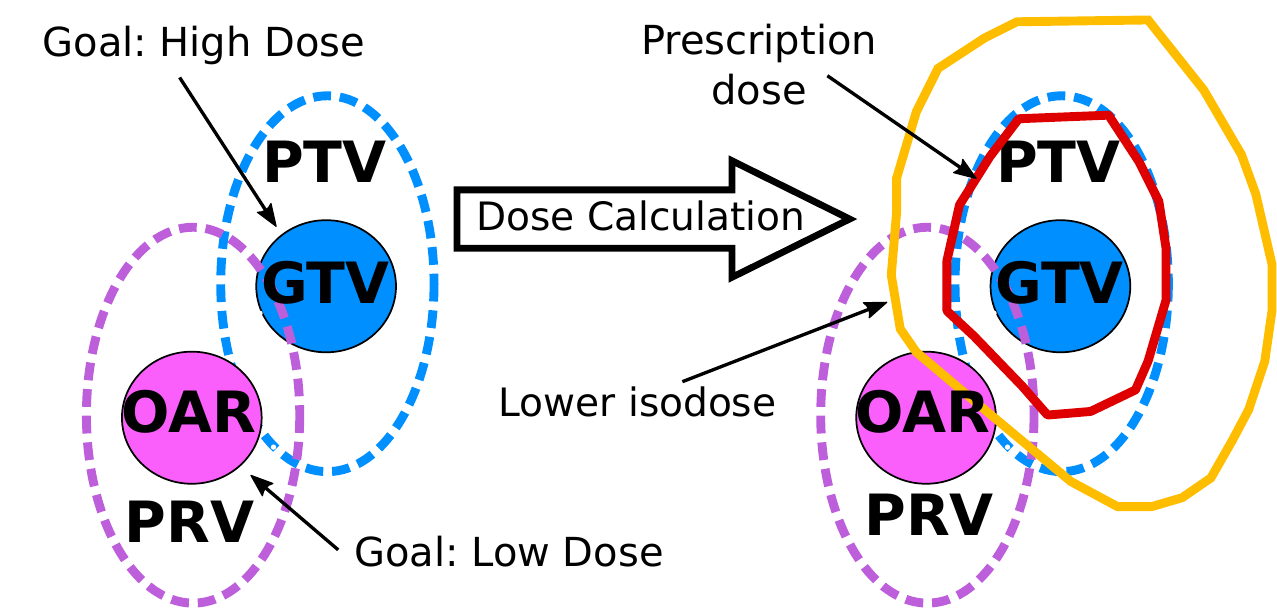}

    \caption{
        Conflicting treatment goals: PTV
        requires high dose, whereas PRV requires low dose. After dose
        calculation the desired treatment goals might not be achieved.
    }
    \label{fig:overlap_problem}
\end{figure}

The process of radiotherapy planning (RT) involves several important steps, such as contouring of target volumes and organs at risk (OARs), and designing the delivery of doses to targets and setting dose constraints for OARs~\cite{Washington2015}.
This process also includes calculating the dose distribution~\cite{Washington2015} and assessing the results before delivery.
However, it is widely acknowledged that the decision-making process is subjective and could benefit from more efficient and reliable assessment tools~\cite{Alfonso2015}.
Some key volume definitions~\cite{icru62} used in RT planning (see Fig.~\ref{fig:overlap_problem}) include the visible gross tumor volume (GTV), clinical target volume (CTV), and planning target volume (PTV).
The CTV and PTV are created by adding margins to account for uncertainties such as tumor extensions and patient positioning. Similarly, the planning risk volume (PRV) is defined for OARs.
Evaluating treatment plans typically involves analyzing the dose volume histogram (DVH) \cite{Alfonso2015}, however, this method has limitations such as the lack of spatial information and the assumption that organ function is evenly distributed within an organ, which may not always be the case\cite{Bentzen2010}.
In this work, we concentrate on the issue of target and OAR volume overlap. Particularly, when PTV and PRV overlap as shown in Fig.~\ref{fig:overlap_problem}, conflicting dose limitations can result in high doses in small parts of OARs.
However, if the overlap between OARs and PTV is minimal, treatment planners might not make a priority to spare these organs more than the standard goal~\cite{Moore2011}.

Our approach combines visual analytics techniques with traditional volume visualization and interaction methods to interactively explore overlapping targets and OARs. This enables a fast, flexible, and visually-driven examination of treatment plans, taking into account relevant dose and physical characteristics, to ensure an optimal outcome for both the target and OARs.

\section{Related Work}
\label{subsec:related_work}

One way to address the issue of sparing healthy tissue during treatment planning is to use a modeling approach that calculates a quality score based on prior knowledge. Another approach is to utilize exploratory visualization techniques that involve the expert knowledge of planners in the decision-making process.

\subsection{Modeling or Knowledge Based Approaches}

A recent review of modeling or knowledge-based approaches can be found in
the surveys by Wang et al.~\cite{Wang2019} and by Ge and Wu~\cite{Ge2019}.
Alfonso et al.~\cite{Alfonso2015} proposed an automated scoring system based on DVHs for comparing multiple tentative plans and providing estimates for PTV coverage and OAR sparing.
Wu et al.~\cite{Wu2009} used the Overlap Volume Histogram~\cite{Kazhdan2009} to guide planners in determining the feasibility of delivering lower doses to OARs in the new plan by identifying patients with similar geometries.
Song et al.~\cite{Song2015} developed a numerical quality indicator using a geometry-dosimetry model to categorize potential plans as optimal or suboptimal, providing patient-specific quality control.
Petit et al.~\cite{Petit2012} proposed a model to evaluate the quality of treatment plans using prior patient data, considering the orientation and distance of OARs to the PTV, resulting in a significant reduction in OAR dose.
A model to guide management of the target volume and OAR overlap was proposed by Mattes et al.~\cite{Mattes2014}. This model accurately guides physicians using the degree of overlap and limits the extent of the overlap region prior to optimization.
Deshpande et al.~\cite{Deshpande2016} proposed a decision support system that takes into account the relationships between the tumor and OARs to find similar cases in a database for estimating acceptable dose distribution.

\subsection{Related Visualization Approaches}
\label{subsec:rw_vis}
A recent overview of visual computing in RT and multi-modality visualization can be found in the surveys by Schlachter et al.~\cite{Schlachter2019a,Schlachter2019b}, Preim et al.~\cite{Preim2016}, and Lawonn et al.~\cite{Lawonn2018}.
Overlap volumes can be treated as spatial sets with specific physical and dose measurements, and thus set visualization techniques can be utilized.
A comprehensive survey of techniques for visualizing set relations is available in Alsallakh et al.~\cite{Alsallakh2015}. Interactive visual analysis of overlaps of arbitrary sets is particularly relevant~\cite{Alsallakh2013}.
Techniques used in other domains, such as analyzing similarities and differences in genomes~\cite{Krzywinski2009} or visualizing adjacency relations~\cite{Holten2006} using chord diagrams, can be applied to overlap relations.
The work by Nunes et al.~\cite{Nunes2014} presents an approach that combines quantitative and spatial views for defining target volumes using MR-Spectroscopy data.

\section{Data, RT Plan Metrics, and Definitions}
\label{sec:design}

\subsection{Data}
\label{subsec:volumetric_data}
The spatial views in our application rely on volumetric data. They typically utilize multi-modal imaging sources such as 4D-PET/CT, segmentations, and dose distributions~\cite{Schlachter2017}. The colors and names of the segmentations are predefined and imported from DICOM-RT. The segmentations are transformed into binary volumes based on the resolution of the planning CT. Both spatial and quantitative views utilize dose distribution data.

\subsection{Important Overlap Relations}
\label{subsec:overlap_relations}
In the following, a structure is defined as spatial set
$S \subseteq \mathbb{R}^3$.
The overlap relation of two structures
$S_i, S_j \subset \mathbb{R}^3$ is defined by
$S_i \sim S_j \iff S_i \cap S_j \neq \emptyset$,
where $i,j \in \{1 \dots N\}$, $i \neq j$, and $N$ is the number of structures.
\label{eq:ov_relation}
If a structure is selected in one of the views, it is denoted as
reference structure $S_r$, where $r \in \{1 \dots N\}$.
In the following $S_i$, $S_j$, $S_r$ are used to denote structures for $i, j, r \in \{1 \dots N\}$ with $i \neq j\neq r$.

\paragraph{Volume-to-Volume Overlaps}
These overlaps are defined by the intersection of delineation
volumes such as the PTV and OAR.
We define an overlap volume as
$OV(S_i, S_j) := S_i \cap S_j$.

\begin{figure}[tb]
    \centering
    \subfloat[]{
        \includegraphics[height=3.55cm]{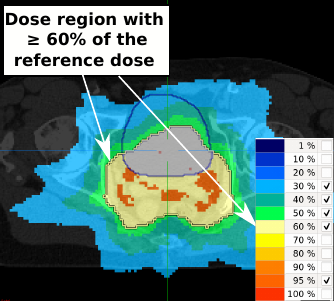}
        \label{fig:define_by_dose_a}
    }
    \subfloat[]{
        \includegraphics[height=3.55cm]{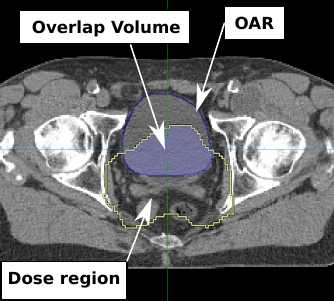}
        \label{fig:define_by_dose_b}
    }
    \caption{Volume-to-Dose Overlaps are defined dose regions $R_d$ \protect\subref{fig:define_by_dose_a} (yellow outline where dose is $\geq d$) and other structures \protect\subref{fig:define_by_dose_b}.}
    \label{fig:define_by_dose}
\end{figure}

\paragraph{Volume-to-Dose Overlaps}
These overlaps are defined between structures $S_i$ and dose
regions $R_d$ (see Fig.~\ref{fig:define_by_dose}) as $OV(S_i, R_d)$.
Given a dose distribution
$f_D: \mathbb{R}^3 \rightarrow \mathbb{R}$
representing the absorbed dose $d$ at position $x\in\mathbb{R}^3$,
a dose region
$ R_d := \{ x \in \mathbb{R}^3 \mid f_D(x) \geq d,~ d \in \mathbb{R} \} $
is defined as the volume where the dose is at least $d$ Gy.

\paragraph{Overlaps by Point Selection}
\label{para:transitive}
The slice views are used to define the set $OV_x$ of all structures, which overlap in the current cross-hair position $x$.
This set is used to display all available overlap information in a table view.
The set is updated when the cross-hair is moved, and individual overlaps can be displayed on-demand.
Given $x \in \mathbb{R}^3$ the set
$ OV_x := \{ S_i \mid S_i \sim S_j \wedge x \in S_i \cap S_j \} $
includes all structures, which overlap each other in the point $x$.

\subsection{Treatment Plan Evaluation Metrics}
\label{subsec:evaluation_metrics}
In the quantitative views, we employ treatment plan evaluation metrics to filter and sort overlaps, and a variety of metrics can be applied~\cite{Moore2012}. The metrics discussed below are suggested as a means to identify overlaps that may warrant lower doses, and can be interchanged without altering the concept of our approach.

For the quantification of dose, we included the minimum $d_{min}$, mean $d_{mean}$, and maximum $d_{max}$ dose value in a structure or overlap volume, and the Homogeneity Index~\cite{Kataria2012} (HI) representing the uniformity of the dose distribution in the volume.
Quantitative metrics for overlap volumes are the absolute volume in \mbox{ml} of the overlap volume $OV(S_i, S_j)$, and the overlap in
percentage relative to the selected structures.
These values provide an insight into whether the volume is significant enough to be investigated further, as for example, a tiny overlap may be deemed insignificant or vice versa.
Of particular interest to us is the difference between the DVHs of the original structure and the overlap DVHs. If the discrepancy is significant, the overlap warrants further examination. The histogram intersection~\cite{Barla2003} is a metric that quantifies the similarity between two distributions. We include the inverse of the histogram intersection in the quantitative views to calculate the dissimilarity between two DVHs
$D_{HI}(S_i,S_j) = 1 -
    \sum\limits_{k=1}^{n} \min (p_i(d_k),p_j(d_k))$,
where $p_{i}$ and $p_{j}$ are the normalized DVHs of structure $S_i$ and $S_j$, and $n$ is the number of bins of the histogram.

\section{Visualization: Views}
\label{sec:visualization}
To gather information about the properties needed to identify potential dose reduction in overlapping volumes, a questionnaire was initially sent to two experienced radiation oncologists. This information was used to develop a prototype, which was then presented to a radiation oncologist and a medical physicist from another institution for feedback and further improvement.

We employ the concept of multiple coordinated views to display metrics related to overlap regions between structures in a treatment plan. The goal is the identification of areas where the dose to an OAR can be reduced without compromising the treatment outcome for the target (see Fig.~\ref{fig:overlap_problem}).
The quantitative views are utilized to depict and encode crucial information about the overlapped volumes and treatment plan evaluation metrics. They are composed of various components, including a chord diagram, parallel coordinates, two distinct table views, and a DVH view.
The spatial views consist of slice views and volume visualization, which are used to display spatial information.

\begin{figure*}[tb]
    \centering
    \subfloat[]{
        \includegraphics[height=5.5cm]{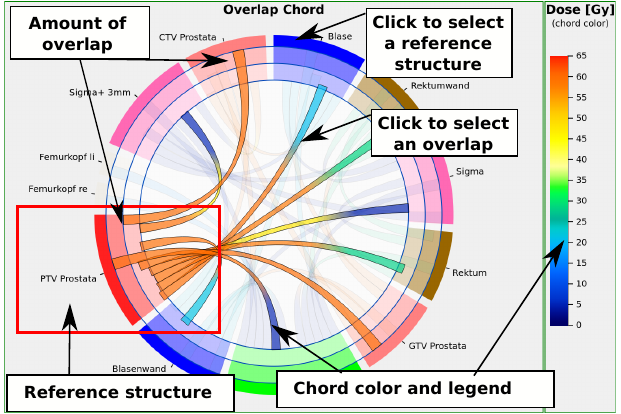}
        \label{fig:chord_a}
    }
    \subfloat[]{
        \includegraphics[height=5.5cm]{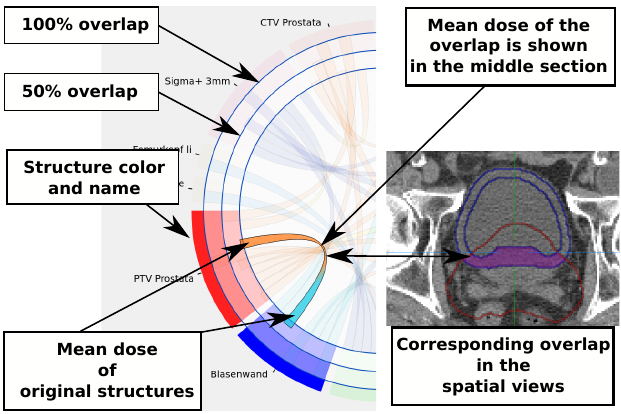}
        \label{fig:chord_b}
    }
    \caption{The chord diagram with a selected reference structure \protect\subref{fig:chord_a}, and the outer ring indicating the amount of overlap \protect\subref{fig:chord_b}. The colors represent the mean dose.
    Here the selected reference structure is the PTV of the prostate \protect\subref{fig:chord_a}, and the overlap with the bladder wall is shown in \protect\subref{fig:chord_b}.
    }
    \label{fig:chord}
\end{figure*}

\subsection{Chord Diagram}
\label{subsec:chord}

The chord diagram provides a compact overview of all overlap relations and helps to easily select the reference structure (see Fig.~\ref{fig:chord_a}).
A chord connection represents an overlap between two volumes, and the color of the chord shows the mean dose value $d_{mean}$ of the overlap volume $OV(S_i, S_j)$ in the middle part, and $S_i$ and $S_j$ on the outer parts.
This allows for comparison of the mean dose values of the original volumes to the overlap.
The ring of the chord diagram has three sections: the outer section represents the structure with its assigned name and color, the inner two sections encode the amount of overlap of the structures in percentage, and the three blue circles represent 0\%, 50\%, and 100\% overlap (see Fig.~\ref{fig:chord_b}).
A structure can be selected as the reference structure by clicking on the outer section, and the portion of arc length allocated to a structure is determined by the number of volumes it overlaps with.

\subsection{Parallel Coordinates View}
\label{subsec:paracoord}

\begin{figure*}[ptb]
    \centering
    \subfloat[]{
        \includegraphics[height=4.8cm]{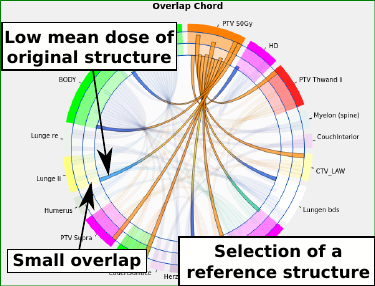}
        \label{fig:step_1_chord}
    }
    \subfloat[]{
        \includegraphics[height=5.3cm]{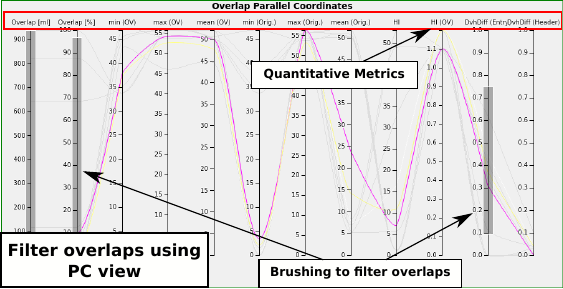}
        \label{fig:step_2_pc}
        \label{fig:paracoord}
    }\\
    \subfloat[]{
        \includegraphics[height=4.8cm]{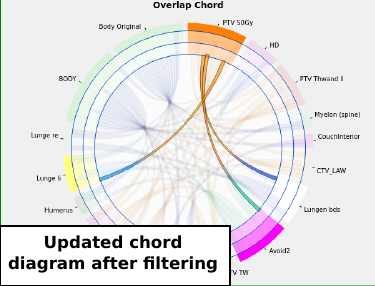}
        \label{fig:step_2_chord}
    }
    \subfloat[]{
        \includegraphics[height=5.1cm]{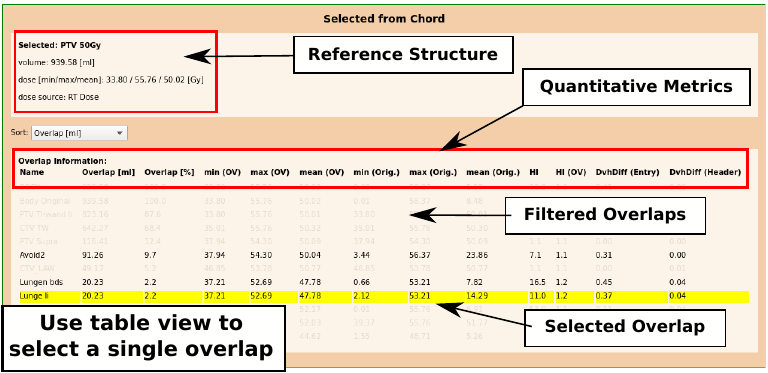}
        \label{fig:step_3_table}
        \label{fig:table_view}
    }\\
    \subfloat[]{
        \includegraphics[height=5cm]{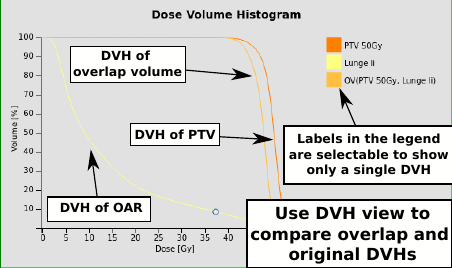}
        \label{fig:step_3_dvh}
        \label{fig:dvh_view}
    }
    \subfloat[]{
        \includegraphics[height=4.9cm]{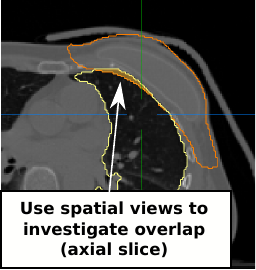}
        \label{fig:step_4_axial}
    }
    \subfloat[]{
        \includegraphics[height=4.9cm]{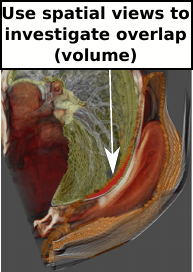}
        \label{fig:step_4_volume}
    }
    \caption{Example exploration starting by selecting a reference structure in the cord diagram \protect\subref{fig:step_1_chord}. The \emph{PTV 50Gy} was selected as reference structure. Filtering in the parallel coordinates \protect\subref{fig:step_2_pc} reduces possible candidates \protect\subref{fig:step_2_chord}, \protect\subref{fig:step_3_table}. Finally, a selection of a single overlap \protect\subref{fig:step_3_table} shows the DVHs \protect\subref{fig:step_3_dvh} and spatial views \protect\subref{fig:step_4_axial}, \protect\subref{fig:step_4_volume} of the corresponding structures alongside the overlap volume.
    }
    \label{fig:example_exploration_I}
\end{figure*}

Parallel coordinates (PC) are a widely used visualization technique for multivariate data, and a well-known visualization for exploratory data analysis~\cite{Heinrich2012a}.
We use a PC view to visualize for each overlap volume $OV( S_i, S_j)$ the corresponding metrics.
The PC was chosen due to its easy-to-understand way of representing multidimensional points (in our case the metrics for each overlap relation), its scalability (easy to add more metrics), and the effective selection of value ranges (brushing) for each metric.
Each dimension in the PC view (see Fig.~\ref{fig:paracoord}) corresponds to metrics calculated for $OV(S_i, S_j)$, including values for $S_i$ and/or $S_j$ depending on whether a reference structure was selected.

If a reference structure $S_r$ is selected in the chord diagram (see Fig.~\ref{fig:chord_a}), each line in the PC view corresponds to a visible chord in the diagram, i.e., all $S_r \sim S_i$ where $i$ corresponds to a visible chord.
In the PC view, the amount of overlap in \mbox{ml} and percentage is displayed in regard to $S_r$.
For $OV(S_r, S_i)$ and $S_i$, the values displayed are $d_{min}$, $d_{mean}$, $d_{max}$ and $HI$.
Furthermore, for $S_r$ and $S_i$ the values
$D_{HI} (S_r, OV(S_r, S_i))$ and $D_{HI} (S_i, OV(S_r, S_i))$ are displayed.
If no reference structure is selected, all overlap volumes $OV(S_i, S_j)$  are displayed in the PC view.

\subsection{Table Views}

We include two table views: TV-I shows all overlaps of a reference structure $S_r$ (see Fig.~\ref{fig:table_view}), TV-II shows all overlaps at a 3D world position.

\paragraph{Reference Structure (TV-I)}
The selected $S_r$ is displayed above the table together with relevant information, such as total volume in \mbox{ml} and dose quantities
(see Fig.~\ref{fig:table_view}).
Each row represents an overlap $S_r \sim S_j$ with the reference structure, and displays metrics for the overlap volume $OV(S_r, S_j)$ and for the structure $S_j$.
The displayed metrics are the same as displayed in the PC view with a selected $S_r$ as described in Section~\ref{subsec:paracoord}.
The table can be sorted, and entries can be selected.

\paragraph{3D World Position (TV-II)}
TV-II depends on the intersection of the slice views $x \in \mathbb{R}^3$ to define the set $OV_x$ (see Section~\ref{para:transitive}).
The values displayed
are the same as in TV-I, but with additional columns. Since there is no $S_r$ defined, values such as $d_{min}$ or volume in \mbox{ml} are displayed for both $S_i$ and $S_j$.

\subsection{Dose Volume Histogram (DVH) View}

DVHs~\cite{Washington2015} offer a valuable compact representation of the 3D dose distribution, widely employed in RT planning.
In our application, the DVH view (see Fig.~\ref{fig:dvh_view}) is dynamically generated based on selections made in other views.
When a single overlap is selected, for example in the chord diagram (see Fig.~\ref{fig:chord_b}), only the DVHs of the corresponding structures are displayed as shown in Fig.~\ref{fig:dvh_view}.
The label in the color legend can be used to select a single DVH and fade the remaining ones.

\subsection{Spatial Views}
\begin{figure}[tb]
    \centering
    \includegraphics[width=\linewidth]{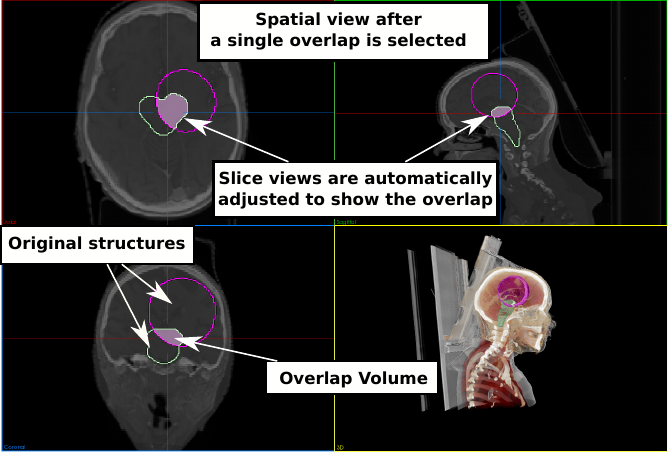}
    \caption{After selecting an overlap in the quantitative views, the spatial views display the original structures and the overlap volume. The overlap is centered in the slice views and displayed as a filled, semi-transparent area.}
    \label{fig:anatomical_overview}
\end{figure}

The spatial views utilize standard techniques for visualizing data in radiation therapy planning~\cite{Schlachter2019b}, with a layout similar to that used in commercial RT planning tools.
They consist of four quadrants, with three displaying slice views and one displaying a 3D view of the data (see Fig.~\ref{fig:anatomical_overview}).
The views are designed to display at least one anatomical image, such as a planning CT, as a reference, and additional images can be added and displayed using image fusion techniques~\cite{Schlachter2017}.
The information displayed in the spatial views is linked to selections made in the quantitative views, with structures such as target volumes, OARs, and overlap volumes being automatically added or removed based on these selections.

\paragraph{Slice Views}
The slice views show image information in coronal, sagittal, and axial
orientation (see Fig.~\ref{fig:anatomical_overview}).
Dose distribution information can be added on demand,
and displayed either as colorwash
or as outline.
For each structure selected for display, its outline is shown in the color assigned to that structure. Additionally, any overlap volumes selected in the quantitative views are displayed as filled, semi-transparent areas, as illustrated in Fig.~\ref{fig:anatomical_overview}. The color of these areas is a mix of the colors of the original structures.

\paragraph{3D View}
The volume visualization of imaging data is based on the approach presented in Schlachter et al.~\cite{Schlachter2017}, and is used for both image fusion and visualization of target and OAR structures.
The implementation also allows for ``cutting'' open the volumes for a closer inspection of the internal information, as shown in Fig.~\ref{fig:anatomical_overview}.
Dose information is displayed as isodose surfaces for specified dose values.
Additionally, it is possible to highlight the surface of the overlap volume in 3D if needed.

\subsection{Implementation}
The implementation uses the MITK~\cite{Wolf2005} platform at its core. The quantitative views are implemented as a plugin in MITK using the D3 library \url{https://d3js.org/} and Qt WebKit \url{https://www.qt.io/} for rendering.
The spatial views are based on the implementation of Schlachter et al.~\cite{Schlachter2017}.

\section{Data Exploration and User Interaction}
\label{sec:interaction}

\begin{figure*}[tbp]
    \centering
    \includegraphics[width=\linewidth]{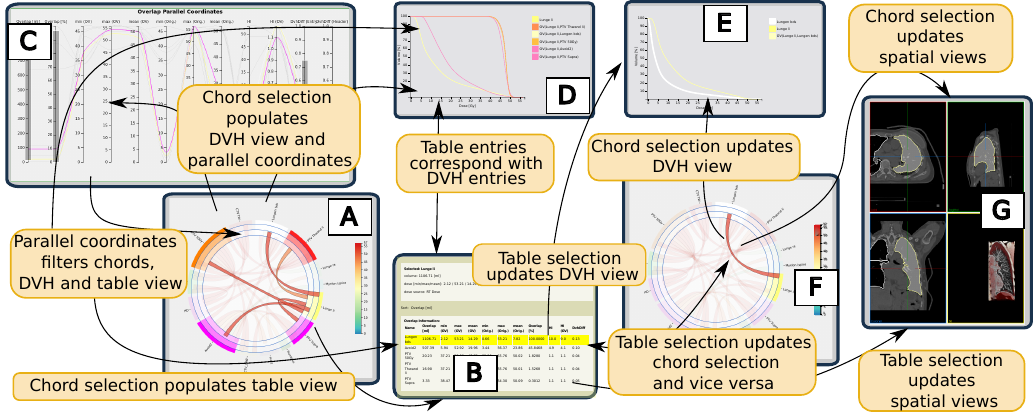}
    \caption{Overview of the interaction between individual quantitative views and the spatial views. Arrow directions indicate which views are updated after an interaction.
    }
    \label{fig:interaction_overview}
\end{figure*}

Before interacting with the views, the underlying data structures must be initialized.
%To compute the quantitative metrics, a dose distribution must be selected.
Optional setup steps include pre-selecting specific dose values for generating dose regions $R_d$ and pre-selecting contours to reduce the data.
The objective is to use the quantitative views to identify potential overlapping areas that could benefit from reduced dose.
The metrics displayed in the different views can be used to narrow down the candidates and examine them in more detail.
A summary of the view connections is shown in Fig.~\ref{fig:interaction_overview}.
The overall approach followed is an overview-first, zoom-and-filter, then details-on-demand method\cite{Shneiderman1996}.

\subsection{Chord Diagram}
The chord diagram is used to select a reference structure as shown in Fig.~\ref{fig:chord_a} and in Fig.~\ref{fig:interaction_overview}~A,
or to select a single chord as shown in Fig.~\ref{fig:chord_b} and Fig.~\ref{fig:interaction_overview}~F.

When a reference structure $S_r$ is selected, the table view TV-I displays all structures that overlap with $S_r$ (see Fig.~\ref{fig:interaction_overview} A to B).
Additionally, the $S_r$ is added to the spatial views and the center of the slice views is set to the center of $S_r$. The parallel coordinates are also updated to only show entries that correspond to the visible chords and entries in TV-I (see Fig.~\ref{fig:interaction_overview} A to C). The DVH view is also updated, displaying only DVHs that correspond to the overlap volumes of the visible chords, including $S_r$ itself (see Fig.~\ref{fig:interaction_overview} A to D).

If a single, outgoing chord of $S_r$ is selected, then the corresponding entry in TV-I is highlighted in yellow (see Fig.~\ref{fig:interaction_overview} F to B), and the DVH view is updated to show only $S_r$, the overlapping structure $S_i$, and the overlap volume $OV(S_r,S_i)$ (see Fig.~\ref{fig:interaction_overview} F to E).
Furthermore, the spatial view is updated to show $S_r$, $S_i$ and $OV(S_r,S_i)$
(see Fig.~\ref{fig:interaction_overview} F to G).
Independently, a single chord can always be selected, but only updates the DVH and spatial view if no $S_r$ is selected.

\subsection{Parallel Coordinates}
If a reference structure $S_r$ is selected the parallel coordinates (PC) view shows all overlaps $S_r \sim S_i$.
Brushing the coordinates, i.e., selecting a value range on the axis (see Fig.~\ref{fig:paracoord}), filters table entries in TV-I, and fades items, which are not selected by the brush (see Fig.~\ref{fig:interaction_overview} C to B).
The same applies to outgoing chords of $S_r$ in the chord diagram (see Fig.~\ref{fig:interaction_overview} C to A), and furthermore the DVH view, which now only shows DVHs of overlaps with the selected criteria (see Fig.~\ref{fig:interaction_overview} C to D).

If no $S_r$ is selected, all overlaps $S_i \sim S_j$ are displayed, i.e., each line in the PC view corresponds to a chord in the chord diagram.
Brushing the coordinates filters the set of overlaps, and updates the chord diagram to show only chords corresponding to the filtered overlaps.
Additionally, the DVH view is updated to show only DVHs of filtered overlaps, including the DVHs of the original structures defining the overlap.

\subsection{Table Views}
After selecting a reference structure in the chord diagram, the table view TV-I can be used to investigate overlaps by selecting one by one.
A selection in the table view (see Fig.~\ref{fig:interaction_overview} B to F) is equivalent to a click on a single chord in the chord diagram.
This also means that a table entry selection updates the chord diagram, the DHV view, and the spatial views.
Overlaps can be deselected and the state before the selection is restored.

A selection in table view TV-II works similarly, but with one key difference. The crosshair in the spatial views is not repositioned. This is to avoid altering the contents of the table. Instead, the selection only updates the zoom and the structures displayed, allowing for a closer examination of overlaps at a specific 3D position.

\subsection{Spatial Views}
The only way to update the quantitative views is by adjusting the position of the slices. This automatically updates the contents of table view TV-II, making all overlap information for the current position readily accessible.

The focus is on updating the spatial views depending on selections made in the quantitative views. For instance, if an overlap is selected in TV-I, an automatic repositioning of the slices to the center of the overlap volume and a zoom depending on the size of the volume is triggered.
Furthermore, only the corresponding structures and overlaps of the selection are displayed (see Fig.~\ref{fig:anatomical_overview}).
There is an option to automatically set isodose values from the selected overlap volume for 3D display, as well as, an optional automated clipping in the 3D view to the bounding box of the selected overlap volume.

\section{Proposed Workflow for RT Plan Evaluation}
\label{sec:workflow}

The proposed workflow begins by selecting a target volume as the reference structure $S_r$ in the chord diagram. This makes all overlapping OARs visible through their outgoing chords, meaning all $S_j$ such that $S_r \sim S_j$. The chord diagram encodes information on the amount of overlap and mean dose values for the volumes $OV( S_r, S_j)$, $S_r$, and $S_j$ (see Fig.~\ref{fig:step_1_chord}).
The PC view can then be used to narrow down the list of candidate volumes (see Fig.~\ref{fig:step_2_pc}). The remaining volumes can be evaluated individually by selecting each entry in the table view TV-I (see Fig.~\ref{fig:step_3_table}). This allows for inspection of the corresponding DVHs in the DVH view (see Fig.~\ref{fig:step_3_dvh}), as well as the anatomy (see Fig.~\ref{fig:step_4_axial} and Fig.~\ref{fig:step_4_volume}) which is automatically brought into focus with the display of $OV( S_r, S_j)$, $S_r$, and $S_j$. Further adjustments can be made in the spatial views if needed.

\subsection{Finding Candidate Overlaps}
\label{subsec:find_candidates}
The process of finding candidate volumes for further analysis is illustrated in Fig.~\ref{fig:example_exploration_I}.
The PTV with dose $\geq 50~\text{Gy}$ is selected in the chord diagram as reference structure $S_r$ (see Fig.~\ref{fig:step_1_chord}).
Now the chord diagram gives an indication that overlaps are either small, but with a low mean dose of the original structure, or either large (up to 100\% overlap) with a similar mean dose as $S_r$.
The small overlaps are the most interesting, as they might have too much dose in the overlap.

Afterwards, the PC view is utilized to filter by overlap in~\mbox{\%} and in~\mbox{ml}, eliminating tiny volumes, and by $D_{HI}$ value (refer to Fig.~\ref{fig:step_2_pc}).
The structures are then narrowed down to three options (see Fig.~\ref{fig:step_2_chord} and Fig.~\ref{fig:step_3_table}).
The TV-I table view can then be utilized to examine all other metrics and choose one overlap as illustrated in Fig.~\ref{fig:step_3_table}; in this case, the left lung is selected for further analysis through the DVH view and the spatial views.

The result after the selection is shown in Fig.~\ref{fig:step_3_dvh}. It is now possible to inspect the DVHs of the overlap and the original structures alongside the anatomy with the highlighted overlap region, as shown for the axial slice in Fig.~\ref{fig:step_4_axial}. The final result can then be refined by adjusting the 3D view, as illustrated in Fig.~\ref{fig:step_4_volume}.

\subsection{Hot Spot Detection in OARs}
\begin{figure}[tb]
    \centering
    \subfloat[]{
        \includegraphics[height=4.2cm]{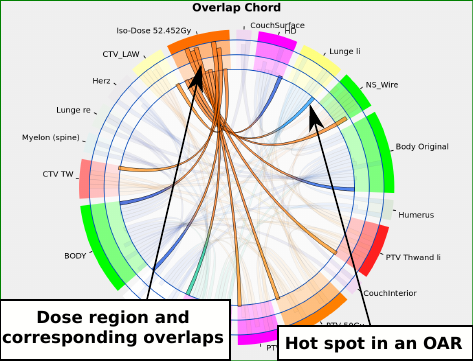}
        \label{fig:hotspot_a}
    }
    \subfloat[]{
        \includegraphics[height=4.2cm]{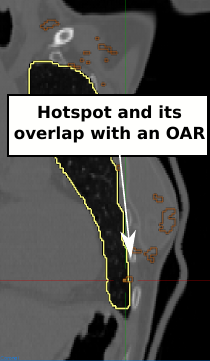}
        \label{fig:hotspot_b}
    }
    \caption{
        Hotspots after defining a dose region with a dose value $d$ corresponding to 93\% of the reference dose.
        All outgoing chords show structures, which have sub-volumes with a dose value $\geq d$ \protect\subref{fig:hotspot_a}. Quantitative measures can be explored using the table view, and further selections update the spatial views \protect\subref{fig:hotspot_b}.}
    \label{fig:hotspot_detection}
\end{figure}

Another application is detecting dose hot spots, as shown in Fig.~\ref{fig:hotspot_detection}. In our application, it is possible to specify dose regions $R_d$ and calculate volume-to-dose overlaps (see Section\ref{subsec:overlap_relations} and Fig.~\ref{fig:define_by_dose}). This is shown in Fig.~\ref{fig:hotspot_a}, where $R_d$ was defined with a dose of 93\% of the reference dose. By selecting $R_d$ as the reference structure in the chord diagram, you can identify potential hot spots by exploring the resulting overlap volumes. As outlined in Section~\ref{subsec:find_candidates}, all volumes with hot spots can be easily located and inspected (see Fig.~\ref{fig:hotspot_b}).

\section{Evaluation and Results}
\label{sec:evaluation}

\begin{table}[htb]
    \caption{Questionnaire per Data Set (Shortened Version)}
    \label{table:questions_ov_short_data}
    %TC:ignore
    % counter
    \newcounter{questionnum}
    \setcounter{questionnum}{0}
    \newenvironment{questionOV}[1][]{\refstepcounter{questionnum}\thequestionnum#1}
    % end counter
    \centering
    \begin{tabular}{l p{0.85\linewidth}}
        \toprule
        Q\questionOV{\label{OV-Q1}} &
        % Question
        Give a final rating of the plan regarding overlap volumes from 1 to 5.
        %(1=\mbox{cannot be improved}--5=\mbox{must be improved})
        \\
        \midrule
        Q\questionOV{\label{OV-Q2}} &
        % Question
        Was your final decision in Q\ref{OV-Q1} based on visualization X?
        \\
        \midrule
        Q\questionOV{\label{OV-Q3}} &
        % Question
        Is visualization X helpful for finding candidate overlaps? (Rating: 1--5)
        %(1=\mbox{not helpful}, 3=\mbox{sometimes helpful}, 5=\mbox{very helpful})
        \\
        \midrule
        Q\questionOV{\label{OV-Q5}} &
        % Question
        %Comprehensibility of visualization X  & (Rating: 1--5)
        Are the visualizations suitable for finding hot spots?
        % in OARs, which need to be reconsidered?
        (Rating: 1--5)
        %Rating from 1--5
        %(1=\mbox{not suitable}, 3=\mbox{sometimes}, 5=\mbox{very suitable})
        \\
        \bottomrule
    \end{tabular}
    %TC:endignore
\end{table}

\begin{table*}[htbp]
    \caption{Questionnaire (Shortened Version)}
    \label{table:questions_ov_short_general}
    % counter
    \newenvironment{questionOV}[1][]{\refstepcounter{questionnum}\thequestionnum#1}
    % end counter
    \centering
    \begin{small}
        \begin{tabular}{l p{0.85\linewidth}}
            \toprule
            Q\questionOV{\label{OV-Q6}}  &
            % Question
            Which of the visualizations are helpful for the problem of identifying overlaps to further reduce the dose?
            (Rating: 1--5)
            %(1=worst, 5=best; same ranking is possible)
            \\
            \midrule
            Q\questionOV{\label{OV-Q7}}  &
            % Question
            Is visualization X suitable for finding candidate overlaps?
            (Rating: 1--5)
            %(1=”not suitable” -- 5=”suitable”)
            \\
            \midrule
            Q\questionOV{\label{OV-Q8}}  &
            % Question
            Intuitiveness: Rate the visualizations for finding candidate overlaps from 1 to 5.
            %(1=incomprehensible--5=intuitive)
            \\
            \midrule
            Q\questionOV{\label{OV-Q9}}  &
            % Question
            Speed: How fast did you find the sub-volumes which could potentially benefit from lower doses with visualization X?
            (Rating: 1--5)
            %(1=snail pace -- 5=very fast)
            \\
            \midrule
            Q\questionOV{\label{OV-Q12}} &
            % Question
            %%%%%%%%%%%%%%%%%%%%% this is maximum length
            Details: Does the visualization of the X help for detailed inspection of candidate volumes?
            (Rating: 1--5)
            % Rating from 1--5
            %(1=\mbox{not helpful}, 3=\mbox{sometimes helpful}, 5=\mbox{very helpful})
            \\
            \midrule
            Q\questionOV{\label{OV-Q13}} &
            % Question
            %Interaction I:
            Does interaction with visualization X give a good spatial overview?
            %Here this means whether using the interaction to synchronize to the Spatial Views help to give a spatial overview, as each selected volume will be presented then in the anatomical view. Rating from 1--5
            (Rating: 1--5)
            %(1=\mbox{not helpful}, 3=\mbox{sometimes helpful}, 5=\mbox{very helpful})
            \\
            \midrule
            Q\questionOV{\label{OV-Q14}} &
            % Question
            %Interaction II:
            Does interaction with the spatial views (reference point selection) help to give an overview of candidate overlaps.
            (Rating: 1--5)
            % Rating from 1--5
            %(1=\mbox{not helpful}, 3=\mbox{sometimes helpful}, 5=\mbox{very helpful})
            \\
            \midrule
            Q\questionOV{\label{OV-Q15}} &
            % Question
            Chord diagram:
            (a) Provides a good overview for candidate overlaps?
            (b) Are the rings good for an overview of the amount of overlap?
            (c) Are the chord colors a good indication for the amount of dose on the original and overlap?
            (d) Is the interaction with the diagram helpful?
            (Rating: 1--5)
            % Rating from 1--5
            %(1=\mbox{not helpful}, 3=\mbox{sometimes helpful}, 5=\mbox{very helpful})
            \\
            \midrule
            Q\questionOV{\label{OV-Q16}} &
            % Question
            Parallel Coordinate: Is brushing helpful to filter for candidate overlaps?
            (Rating: 1--5)
            % Rating from 1--5
            %(1=\mbox{not helpful}, 3=\mbox{sometimes helpful}, 5=\mbox{very helpful})
            \\
            \midrule
            Q\questionOV{\label{OV-Q17}} &
            % Question
            %Table view
            TV-I:
            (a) Is the table view helpful for finding candidate overlaps?
            (b) Is the interaction with the table view good for investigating candidate overlaps?
            (Rating: 1--5)
            % Rating from 1--5
            %(1=\mbox{not helpful}, 3=\mbox{sometimes helpful}, 5=\mbox{very helpful})
            \\
            \midrule
            Q\questionOV{\label{OV-Q18}} &
            % Question
            %Table view
            TV-II: Same questions as Q\ref{OV-Q17}
            \\
            \midrule
            Q\questionOV{\label{OV-Q19}} &
            % Question
            DVH view: Is it helpful to show the originals together with overlap volumes?
            (Rating: 1--5)
            % Rating from 1--5
            %(1=\mbox{not helpful}, 3=\mbox{sometimes helpful}, 5=\mbox{very helpful})
            \\
            \midrule
            Q\questionOV{\label{OV-Q21}} &
            % Question
            %Use cases:
            Is adding dose regions helpful for investigating overlap volumes?
            \\
            \midrule
            Q\questionOV{\label{OV-Q22}} &
            % Question
            Can you do the same with the standard software, i.e., finding overlap volumes with too high doses or hot spots?
            % yes no sometimes
            Would it take longer?
            \\
            \midrule
            Q\questionOV{\label{OV-Q23}} &
            % Question
            Do you see potential to integrate the visualizations in the radiotherapy planning workflow?
            % yes    Please specify where
            % no   Please specify why not
            \\
            \midrule
            Q\questionOV{\label{OV-Q24}} &
            % Question
            What are missing features with regard to finding sub-volumes of OARs which are receiving too much dose?
            \\
            \bottomrule
        \end{tabular}
    \end{small}
    %\end{footnotesize}
\end{table*}

We conducted a user study with three experienced domain experts: a radiation oncologist, a medical physicist, and a radiation therapist, each with over 10 years of experience.
They were asked to use the software to evaluate dose plans for five data sets and complete a questionnaire. Each data set represented a different type of cancer (see Table~\ref{tab:question1}).
The experts were tasked with examining the cases, finding potential candidates for dose reduction, and locating dose hot spots.
Questions Q\ref{OV-Q1}--Q\ref{OV-Q5} were answered for each data set (see Table~\ref{table:questions_ov_short_data}) and questions Q\ref{OV-Q6}--Q\ref{OV-Q24} were answered after completing the tasks
(see Table~\ref{table:questions_ov_short_general}).
Responses were given as a rating from 1 to 5, where a higher score indicates better performance, or as ``yes''/``sometimes''/``no'', or as free text. Users could provide additional comments for each question.

For Q\ref{OV-Q1}, users were asked to rate the plan with regard to overlap volumes on a scale of 1 (not improvable) to 5 (must be improved). A rating of 3 indicates room for improvement but not crucial. Results showed that in 4 of 5 cases, at least 2 users agreed on a rating (see Table~\ref{tab:question1}). However, in three cases there was at least one user who disagreed with the others in terms of acceptance.
\begin{table}[tb]
    \caption{Final rating of the plan with regard to overlap volumes.}
    \label{tab:question1}
    %TC:ignore
    \centering
    \begin{small}
        \begin{tabular}{lccccc}
            \toprule
                         &    &    &    & Same   & Agree on   \\
            Data Set     & U1 & U2 & U3 & Rating & Acceptance \\
            \midrule
            Abdomen      & 3  & 2  & 5  & 0      & no         \\
            Head \& Neck & 2  & 2  & 5  & 2      & no         \\
            Prostate     & 5  & 2  & 5  & 2      & no         \\
            Brain        & 5  & 4  & 5  & 2      & yes        \\
            Thorax       & 1  & 1  & 2  & 2      & yes        \\
            \bottomrule
        \end{tabular}
    \end{small}
    %TC:endignore
\end{table}
They were asked in Q\ref{OV-Q2} on which of the visualizations their final decision was based on (see Table~\ref{tab:results_Q2}).
In \textbf{93}\% the chord diagram is used for the decision, followed by the TV-I and spatial views with \textbf{87}\%, and the DVH view with \textbf{67}\%.
\begin{table*}[tb]
    \caption{User ratings for Q\ref{OV-Q2} (in percentages), Q\ref{OV-Q3}, Q\ref{OV-Q5}--Q\ref{OV-Q9}, Q\ref{OV-Q12} and Q\ref{OV-Q13}. Q\ref{OV-Q2}, Q\ref{OV-Q3} and Q\ref{OV-Q5} are aggregated over all test data sets.}
    \label{tab:results_Q_combined}
    \label{tab:results_Q2}
    \label{fig:results_II}
    \label{fig:results_II_1}
    \label{fig:results_II_2}
    %TC:ignore
    \centering
    \begin{tabular}{lrrrrrrrrrr}
        \toprule
                             & \multicolumn{2}{c}{Q\ref{OV-Q2}} & Q\ref{OV-Q3} & Q\ref{OV-Q5}                 & Q\ref{OV-Q6} & Q\ref{OV-Q7} & Q\ref{OV-Q8} & Q\ref{OV-Q9} & Q\ref{OV-Q12} & Q\ref{OV-Q13} \\
        \midrule
                             & | yes                            & no |         & \multicolumn{8}{|c|}{Rating}                                                                                             \\
        \midrule
        Chord Diagram        & \textbf{93}                      & 7
                             & 4.6
                             & 4.13
                             & 4.33
                             & 4.33
                             & 3.67
                             & 4
                             & 3
                             & 4.67
        \\
        Parallel Coordinates & 20                               & \textbf{80}
                             & 1.6
                             & 1.8
                             & 1.67
                             & 2
                             & 1.67
                             & 1.67
                             & 2.33
                             & 2
        \\
        Table View I         & \textbf{87}                      & 13
                             & 4.47
                             & 4.27
                             & 4
                             & 3.33
                             & 4
                             & 4
                             & 4.67
                             & 4
        \\
        Table View II        & 0                                & \textbf{100}
                             & 1.27
                             & 1.53
                             & 1.33
                             & 1.33
                             & 3
                             & 1.67
                             & 2.33
                             & 2.33
        \\
        DVH                  & \textbf{67}                      & 33
                             & 3.93
                             & 3.4
                             & 3.67
                             & 3.67
                             & 4.33
                             & 3
                             & 3.67
                             & 2.33
        \\
        Spatial Views        & \textbf{87}                      & 13
                             & 4.72
                             & 4.53
                             & 5
                             & 5
                             & 5
                             & 5
                             & 4.67
                             & 5
        \\
        \bottomrule
    \end{tabular}
    %TC:endignore
\end{table*}
In Q\ref{OV-Q3} we asked which visualizations are helpful for solving the task, which also shows that these views are helpful, whereas TV-II and parallel coordinates are rated as not helpful (see Fig.~\ref{fig:results_II}).
In Q\ref{OV-Q5} we asked if the visualizations are suitable for finding hot spots in OARs. The results show that the spatial views, TV-I and chord diagram are the most suitable for this task (see Fig.~\ref{fig:results_II}).
Q\ref{OV-Q6} to Q\ref{OV-Q9} assess the helpfulness in detecting overlaps for dose reduction (Q\ref{OV-Q6}), the ability to identify potential overlap candidates for closer examination (Q\ref{OV-Q7}), the visualizations' intuitiveness (Q\ref{OV-Q8}), and the speed of candidate identification (Q\ref{OV-Q9}). Results showed that the spatial views were rated highest. The chord diagram, TV-I, and DVH view received ratings greater than \textbf{3}, while parallel coordinates and TV-II were rated at or below \textbf{3} (see Fig.~\ref{fig:results_II}).

In Q\ref{OV-Q12}, we asked which visualizations aid in thorough examination of candidate volumes. Results indicate that spatial views and TV-I are deemed most helpful, with DVH view coming next (see Fig.~\ref{fig:results_II}), which aligns with our proposed workflow.
In Q\ref{OV-Q13}, we inquired if utilizing the interaction in quantitative views to sync with spatial views enhances spatial understanding. Results show that the chord diagram and TV-I are considered useful, in addition to the spatial views (see Fig.~\ref{fig:results_II}). Other views received low ratings.
In Q\ref{OV-Q14}, the interaction that updates TV-II received a rating of \textbf{2.33} and was deemed unhelpful.

Q\ref{OV-Q15} covers the chord diagram (see Fig.~\ref{fig:results_III}). In terms of overview of candidate overlaps and the interaction it is perceived as helpful. The outer rings are seen as occasionally useful, while the chord colors received lower ratings. One suggestion was to use maximum instead of mean dose for coloring.
In Q\ref{OV-Q16}, the effectiveness of brushing parallel coordinates for filtering candidate overlaps was questioned, but rated as unhelpful with a score of \textbf{2.0}.
In Q\ref{OV-Q17} and Q\ref{OV-Q18}, the survey assessed the usefulness of the table views (TV-I and TV-II) for locating candidates and investigating them. Results showed that TV-I was sometimes helpful for finding candidates and helpful for investigating them (see Fig.~\ref{fig:results_III}). However, both aspects were rated as unhelpful for TV-II.
In Q\ref{OV-Q19} users were asked if it is helpful in the DVH view to show the original volumes together with the overlap volumes, and the results were \textbf{33}\% ``yes'' and \textbf{67}\% ``sometimes''.
\begin{table}[tb]
    \caption{User ratings for Q\ref{OV-Q15}, Q\ref{OV-Q17} and Q\ref{OV-Q18}.}
    \label{fig:results_III}
    %TC:ignore
    \centering
    \begin{tabular}{lrrrrr}
        \toprule
                      &               & (a)  & (b)  & (c)  & (d) \\
        \midrule
        Chord Diagram & Q\ref{OV-Q15}
                      & 4.67          & 3.33 & 2.33 & 4.33
        \\
        Table View I  & Q\ref{OV-Q17}
                      & 3.33          & 4    & --   & --
        \\
        Table View II & Q\ref{OV-Q18}
                      & 1.67          & 2.33 & --   & --
        \\
        \bottomrule
    \end{tabular}
    %TC:endignore
\end{table}
In Q\ref{OV-Q21} we asked if adding dose regions is also helpful for investigating overlap volumes and not just for investigating hot spots. This is regarded as very helpful, especially if there is no overlap with another structure to identify dose regions of interest in OARs.
In Q\ref{OV-Q22} we asked if it is possible to do the same with the standard software, and if it would take longer, to explore all overlapping structures.
The answers suggest that it is feasible, but will take significantly more time due to the manual creation of overlaps.
In Q\ref{OV-Q23} we asked about integrating the visualizations in the RT planning process and all participants agreed it was possible and suggested integration during plan creation and optimization, and plan discussion.
We asked in Q\ref{OV-Q24} which features are missing.
One suggestion is to use logical operations for structures to eliminate, for example, the PTV from the body structure to identify hot spots outside the PTV.
Another recommendation is to create a new reference structure from the table view TV-I
without having to go back to the chord diagram.

\section{Discussion}
\label{sec:discussion}
In general the results are promising.
The chord diagram was well perceived in many aspects, most notably to gain an overview.
Also, table view TV-I was generally perceived as very helpful.
However, a user pointed out that it would be beneficial to have information about structures in the table view even if no reference is selected in the chord diagram.
The DVH view was considered helpful, particularly when used with TV-I, to examine overlaps in detail. Selecting a structure in TV-I gives insight into the trade-off between PTV dose coverage and sparing OARs. The display of original volumes alongside overlap volumes was also viewed as helpful. A user commented that the view allows one to determine whether to prioritize PTV dose coverage or OAR sparing, and the dose the spared PTV receives, but acknowledged that it takes time to get used to, as it differs from the current evaluation process.
The parallel coordinates were added as a means to reduce information and therefore possible candidates.
However, it was barely used for that purpose. It was perceived as too complicated, although the idea of filtering was appreciated.
Automatic updates of spatial views based on selections in quantitative views improves focus on a single reference structure for further investigation and is seen as useful in many ways. The TV-II table view was not well received because it requires manual scanning of the image with limited information displayed, to locate a potential candidate.

The three experts were of different professions, which is possibly biasing their view of the cases and the software. This limitation should be considered when interpreting their responses.

\section{Conclusion and Outlook}
\label{sec:conclusion}
We present a novel way of exploring radiotherapy plans with a focus on overlap volumes. We demonstrate that the approach can also be used for hot spot detection.
Our evaluation shows the validity of our approach.

Volume overlaps with PET-based regions for dose boosting~\cite{Even2015}, e.g. dose painting in non-small cell lung cancer, could be another possible application. Volumes based on the standard uptake value (SUV) could be added and used similar to dose regions.
This could also be extended to show multiple overlaps, e.g., the PTV, the SUV volume, and the dose region overlap. Similarly, overlaps with dose regions using past and current dose distributions could be used for re-planning, and help planners to decide where further reduction might be advisable due to past irradiation.

%TC:ignore

\section*{Acknowledgements}
The authors would like to thank Dr. Tanja Schimek-Jasch and Dr. Dr. Oliver Oehlke for filling out the early stage questionnaire used to develop the first prototype.
This paper was partly written in collaboration with the VRVis
Competence Center. VRVis is fund\-ed by BMK, BMAW, Styria, SFG, Tyrol and Vienna Business Agency in the scope of COMET--Competence Centers for
Excellent Technologies (879730), which is managed by FFG.
%TC:endignore

%TC:ignore
\section*{Conflict of interest}

The authors declare that they have no conflict of interest.
%TC:endignore

% \appendix
% \section{Additional Figures}
% Appendix sections are coded under \verb+\appendix+.

% \verb+\printcredits+ command is used after appendix sections to list
% author credit taxonomy contribution roles tagged using \verb+\credit+
% in frontmatter.


%\bibliographystyle{cas-model2-names}

%\bibliography{cas-refs}

%-----------------------------------------------------------------
% bibliography
% ijcars
% needs \usepackage[numbers]{natbib}
\bibliographystyle{unsrtnat}

% \IfFileExists{paper.bib}{
%     %%use following if all content of bibtex file should be shown
%     %\nocite{*}
%     \bibliography{paper.bib}
%     }{
                   
%}


\end{document}